\documentclass[amsmath]{JFM-FLM_Au}

\lefttitle{Teixeira et al.}
\righttitle{Journal of Fluid Mechanics}

\title{Integral constraints on the linear instability of stratified flow with planar shear as a function of the shear angle}

\author{Miguel A. C. Teixeira\aff{1}, Mohamed Foudad\aff{2} \and Paul. D. Williams\aff{2}}
\affiliation{\aff{1}IDMEC, Instituto Superior T\'ecnico, University of Lisbon, Lisbon, Portugal \aff{2}Department of Meteorology, University of Reading, Reading, UK}

\corresau{Miguel A. C. Teixeira, miguel.cortez.teixeira@tecnico.ulisboa.pt}

\begin{document}
\maketitle

\begin{abstract}
Integral constraints on the inviscid linear instability of stably stratified parallel flow with planar shear at an arbitrary angle to the vertical are derived using the analytical approach of Miles and Howard, focusing on perturbations with two-dimensional (2D) spatial symmetry.  The resulting formulation reproduces the Miles-Howard stability condition for vertical shear, but yields no stability condition for non-vertical shear, as expected. A new expression is obtained for the upper bound of the instability growth rate as a function of the shear angle, extending Howard's result and showing consistency with published numerical results for a 2D Bickley jet. Additionally, new bounds on the complex angular frequency of the instability are derived as a function of the shear angle, which reduce to Kochar and Jain's semi-ellipse theorem for vertical shear and to Howard's semi-circle theorem for horizontal shear, having a more complicated form in intermediate cases.
\end{abstract}



\section{Introduction}
\label{sec:intro}
The stability of stratified shear flows is a fundamental problem of wide-ranging relevance in fluid mechanics, having implications for the transition of a flow to turbulence, the dynamics of internal gravity waves, and various applications, such as the generation of clear-air turbulence in the atmosphere and mixing in the oceans or in industrial processes.

The most basic aspects of the linear stability of parallel shear flow go back to \cite{Rayleigh_1880}, who showed that in inviscid non-stratified conditions the flow may be unstable to exponentially-growing normal modes only if an inflection point exists in the mean velocity profile. \cite{Miles_1961}, using a similar approach, showed that for inviscid stratified horizontal flow with planar vertical shear, stability is ensured if the Richardson number ($Ri$) exceeds 1/4. The derivation of this result was considerably simplified by \cite{Howard_1961}, who also derived an upper bound for the growth rate of the instability, and bounds on the magnitude of the complex phase speed of unstable flow perturbations in terms of the range of mean flow velocities, in what is usually known as the semi-circle theorem. While the condition $Ri>1/4$ is sufficient for linear flow stability, and therefore $Ri<1/4$ is a necessary condition for instability, there is no known sufficient condition for instability in general. \cite{Collyer_1970} and \cite{Hazel_1972}, followed by several other authors, confirmed the theoretical predictions of \cite{Miles_1961} and \cite{Howard_1961} through analytical, numerical and experimental studies of stratified shear flow focusing on the instability conditions for specific velocity profiles. Subsequently, \cite{Kochar_Jain_1979} presented an extension of Howard's semi-circle theorem, known as the semi-ellipse theorem, which defines a tighter bound on the complex phase speed of unstable perturbations in planar flow with vertical shear.

The problem of the instability of stratified horizontal flows with horizontal or otherwise non-vertical shear has received comparatively less attention. An early treatment of non-planar shear with both horizontal and vertical components was made by \cite{Eckart_1963}, who extended Howard's semi-circle theorem to adiabatic jets using a Lagrangian displacement formulation of the equations of motion.
Later, \cite{Blumen_1975} derived a generalized version of the semi-circle theorem for generic parallel flows with both horizontal and vertical shear, and also studied the linear stability of such flows for specific velocity profiles. 
Using Eckart's Lagrangian displacement formulation of the equations of motion, \cite{Fung_1986} extended the semi-ellipse theorem of \cite{Kochar_Jain_1979} to stratified parallel flows with both horizontal and vertical shear, and also derived generalized stability conditions as an extension of the Miles-Howard theorem. 

With increasing computational power, the instability of stratified flows with horizontal and/or vertical shear has received renewed attention more recently. \cite{Deloncle_etal_2007} numerically solved the eigenvalue--eigenfunction problem for inviscid stratified flow with a specific horizontal shear profile, corroborating the absence of a stability condition in that case. Using similar methods, \cite{Candelier_etal_2011} analysed the stability of stratified flows with planar shear at an arbitrary angle to the vertical, showing that they may be linearly unstable for all Richardson numbers, except in the case of vertical shear, and quantifying the dependence of the instability growth rate on the stratification and shear angle. \cite{Facchini_etal_2018} and \cite{LeGal_etal_2021} investigated the stability of viscous stratified flows with horizontal shear, both numerically and experimentally. They found that for finite Reynolds number ($Re$) the flow may become stable, with the critical $Ri$ below which instability occurs depending on $Re$. 

A number of studies have extended the Miles-Howard stability theory, semi-circle and semi-ellipse theorems in the context of magnetohydrodynamics, beginning with \cite{Howard_Gupta_1962}, as reviewed recently by \cite{Deguchi_2021}.  Although the present study does not focus on magnetic effects, it must be recognized that Deguchi's study provides one of the most general treatments to date of the inviscid stability of stratified shear flows with or without those effects, in the vein of Miles-Howard and Kochar-Jain, among other original results. In particular, it clarifies the condition derived by \cite{Fung_1986} for the stability of flows with both horizontal and vertical shear, showing that it reduces to the Miles-Howard condition for vertical shear, and cannot be satisfied for non-vertical shear. \cite{Deguchi_2021} also presents extensions of the semi-circle and semi-ellipse theorems that further restrict the bounds on the complex phase velocity of unstable perturbations.

The use of Direct Numerical Simulations (DNS) to investigate the growth of small flow perturbations in linear instability is limited by numerical errors resulting from the finite resolution of the simulations, which constrain the range of achievable $Re$. \cite{Holt_etal_1992}, \cite{Jacobitz_Sarkar_1998} and \cite{Jacobitz_2002} used DNS to study the related problem of turbulent kinetic energy (TKE) growth in fully turbulent flows. They found that the TKE growth rate and the threshold $Ri$ for instability (understood as TKE increase) are higher for horizontal shear than for vertical shear in their intrinsically viscous simulations. This is consistent with the behaviour found in the linear instability analyses of \cite{Facchini_etal_2018} and \cite{LeGal_etal_2021}. However, inviscid theory still provides an essential benchmark for understanding viscous models, where the representation of friction is necessarily imperfect. 
 
Although the Miles-Howard theorem and the semi-circle and semi-ellipse theorems have been extended to flows with both horizontal and vertical shear, the corresponding integral constraints on the inviscid linear instability for planar shear at an arbitrary angle to the vertical have not, to our knowledge, been investigated as functions of that angle. This is the flow configuration addressed numerically by \cite{Candelier_etal_2011}. The present paper aims to fill this gap, following the approach of \cite{Miles_1961}, \cite{Howard_1961} and \cite{Kochar_Jain_1979}. In particular, we derive the dependence of the stability condition, instability growth rate, and complex angular frequency of unstable modes with two-dimensional (2D) symmetry on the shear angle.

Section 2 provides a derivation of the results, section 3 illustrates their behaviour as a function of input parameters, which is discussed, and section 4 summarizes the main conclusions.

\section{Integral constraints for planar shear at an arbitrary angle to the vertical}

\subsection{Basic equations}

We start from the inviscid, adiabatic, linearised equations of motion under the Boussinesq approximation. The adopted coordinate system is inclined at an angle $\theta$ to the vertical, and the mean flow (in hydrostatic balance) is assumed to be aligned in the $x$ direction, with mean shear only in the $z$ direction. This allows the planar shear to range from purely vertical to purely horizontal, depending on $\theta$:
\begingroup
\allowdisplaybreaks
\begin{align}
&\frac{\partial u}{\partial t} + U \frac{\partial u}{\partial x} + w \frac{dU}{dz} = - \frac{1}{\rho_0} \frac{\partial p}{\partial x}, \label{eqmomu} \\
&\frac{\partial v}{\partial t} + U \frac{\partial v}{\partial x} = -\frac{1}{\rho_0} \frac{\partial p}{\partial y} + b \sin \theta, \label{eqmomv} \\
&\frac{\partial w}{\partial t} + U \frac{\partial w}{\partial x} = - \frac{1}{\rho_0} \frac{\partial p}{\partial z} + b \cos \theta, \label{eqmomw} \\
&\frac{\partial b}{\partial t} + U \frac{\partial b}{\partial x} + N^2 w \cos \theta + N^2 v \sin \theta = 0, \label{eqbuoy} \\
&\frac{\partial u}{\partial x} + \frac{\partial v}{\partial y} + \frac{\partial w}{\partial z} = 0. \label{eqmass}
\end{align}
\endgroup
Here, $U(z)$ is the (horizontal) mean flow velocity, $(u,v,w)$ is the three-dimensional (3D) velocity perturbation, $p$ and $b$ are the pressure and buoyancy perturbations, and $N^2$ and $\rho_0$ are the mean static stability and a constant reference density, respectively. The fluid is assumed to be stably stratified, i.e., $N^2>0$. $\theta$ is the angle between the shear and the stratification (which acts in the vertical). $\theta=0$ corresponds to the usual case of aligned shear and stratification (both in the vertical) and $\theta=\pi/2$ corresponds to horizontal shear. The equation set (\ref{eqmomu})-(\ref{eqmass}) is equivalent to equations (2.3a)-(2.3e) of \cite{Candelier_etal_2011}, but with the $y$ axis having a component pointing upwards instead of downwards when $\theta \neq 0$.

Assuming normal-mode solutions of the form $\phi= \hat{\phi} \exp[ i (k x + l y - \omega t)]$ for all perturbation variables, where $(k,l)$ is the wavenumber vector in the $(x,y)$ plane and $\omega$ is the angular frequency, (\ref{eqmomu})-(\ref{eqmass}) can be combined, reducing to a single equation for the amplitude of the velocity perturbation parallel to the shear direction, $\hat{w}$. For two-dimensional (2D) perturbations ($l=0$), that equation takes the form
\begin{equation}
\hat{w}'' + \left\{ k^2 \left[ \frac{N^2 \cos^2 \theta}{(U k - \omega)^2 - N^2 \sin^2 \theta} - 1 \right] -\frac{U'' k}{U k - \omega} \right\} \hat{w} = 0, \label{tgeq}
\end{equation}
where primes denote differentiation with respect to $z$. This is an extension of the Taylor--Goldstein equation for planar shear at an angle to the vertical, which reduces to the standard version when $\theta=0$. With $l=0$, the terms $\partial p/\partial y$ in (\ref{eqmomv}) and $\partial v/\partial y$ in (\ref{eqmass}) vanish, and (\ref{tgeq}) is simplified very substantially compared to the form it would take for $l \neq 0$. This is not, however, equivalent to assuming $v=0$: the velocity perturbations remain 3D, but with a 2D spatial structure. Although Squire's theorem \citep{Squire_1933} is not applicable to this case, this assumption is justified by the perceived fact that the perturbations with the largest growth rate have $l=0$ (i.e., are 2D) for various relevant types of shear flow \citep{Deloncle_etal_2007,Candelier_etal_2011}. Because $z$ is the direction of inhomogeneity for an arbitrary mean velocity profile $U(z)$ and arbitrary angle $\theta$ of the mean shear with the vertical, these assumptions also require $N^2$ to be constant, for consistency.

\subsection{Stability condition}

Following \cite{Howard_1961}, by introducing the change of variable $\hat{w}=(U k - \omega)^{1/2} G$, (\ref{tgeq}) becomes
\begin{equation}
\left[ (Uk - \omega) G' \right]' + \left\{ k^2 (Uk - \omega) \left[ \frac{N^2 \cos^2 \theta}{(Uk- \omega)^2 - N^2 \sin^2 \theta} - 1 \right] - \frac{1}{4} \frac{(U'k)^2}{U k - \omega} - \frac{U'' k}{2} \right\} G = 0. \label{eqg}
\end{equation}
This equation can then be multiplied by the complex conjugate of $G$ ($G^*$) and integrated between two points in $z$, say $z_1$ and $z_2$ at which $G(z)=0$. Since $\hat{w}$ is the amplitude of the perturbation velocity component along $z$, this may be viewed as some form of rigid-lid boundary condition, but as \cite{Howard_1961} originally pointed out, these rigid lids may be moved to as far away as required. This integration yields
\begin{align}
 &\int_{z_1}^{z_2} \left( (U k - \omega) |G'|^2 + \left\{ k^2 (Uk-\omega) \left[ 1 - \frac{N^2 \cos^2 \theta}{(Uk-\omega)^2 - N^2 \sin^2 \theta} \right] \right. \right. \nonumber \\ 
 &\left. \left. + \frac{1}{4} \frac{(U'k)^2}{Uk-\omega} + \frac{1}{2} U'' k \right\} |G|^2 \right) dz = 0, \label{inteqg}
\end{align}
where the term involving $|G'|^2$ results from integration by parts and uses the boundary conditions $G(z_1)=G(z_2)=0$, and from the definitions $|G|^2 = G^* G$ and $|G'|^2 = G'^* G'$. It is convenient to rewrite (\ref{inteqg}) as
\begin{align}
 &\int_{z_1}^{z_2} (Uk-\omega) \left( |G'|^2 + k^2 |G|^2 \right) dz + \frac{1}{2} k \int_{z_1}^{z_2} U'' |G|^2 dz \nonumber \\ 
 &+ k^2 \int_{z_1}^{z_2} (Uk-\omega) \left[ \frac{1}{4} \frac{U'^2}{(Uk-\omega)^2} - \frac{N^2 \cos^2 \theta}{(Uk-\omega)^2 - N^2 \sin^2 \theta} \right] |G|^2 dz = 0. \label{eqintg2}
\end{align}

It should be noted that the angular frequency $\omega$ defining the temporal dependence of the flow perturbations is in general complex: $\omega = \omega_r + i \omega_i$, with real part $\omega_r$ and imaginary part $\omega_i$. The phase speed of the perturbations is $c_r = \omega_r/k$ and their growth rate is $\omega_i$.
Then, the imaginary part of (\ref{eqintg2}) yields
\begin{align}
&\omega_i \left( \int_{z_1}^{z_2} \left( |G'|^2 + k^2 |G|^2 \right) dz \right. \nonumber \\
&\left. + k^2 \int_{z_1}^{z_2} \left\{ \frac{N^2 \cos^2 \theta \left[ (Uk-\omega_r)^2 + \omega_i^2 + N^2 \sin^2 \theta \right]}{\left[(U k-\omega_r)^2 - \omega_i^2 - N^2 \sin^2 \theta \right]^2 + 4 \omega_i^2 (Uk-\omega_r)^2} - \frac{1}{4} \frac{U'^2}{|U k - \omega |^2} \right\} |G|^2 dz \right) \nonumber \\
&= 0, \label{condim}
\end{align}
where $|Uk - \omega|^2=(Uk-\omega_r)^2+\omega_i^2$.
If the expression in the outer round brackets in (\ref{condim}) is positive, then $\omega_i=0$, i.e., no linear instability can occur. Since the first integral is always positive, this requires the second integral to also be positive, which is the case if its integrand is positive everywhere. The integrand in the second integral is positive when
\begin{equation}
\frac{N^2 \cos^2 \theta \left[ (Uk-\omega_r)^2 + \omega_i^2 + N^2 \sin^2 \theta \right]}{\left[ (Uk-\omega_r)^2- \omega_i^2 - N^2 \sin^2 \theta \right]^2 + 4 \omega_i^2 (Uk - \omega_r)^2} > \frac{1}{4} \frac{U'^2}{(Uk-\omega_r)^2 + \omega_i^2}, \label{stabcond}
\end{equation}
for any $(Uk-\omega_r)$, because this is the only term that depends on $z$, via $U(z)$. The condition expressed by (\ref{stabcond}) is then satisfied if
\begin{equation}
   4 \frac{N^2 \cos^2 \theta}{U'^2_{max}} > {\rm Max} \left\{ \frac{\left[ (Uk-\omega_r)^2-\omega_i^2- N^2 \sin^2 \theta \right]^2 + 4 \omega_i^2 (Uk-\omega_r)^2}{\left[ (Uk-\omega_r)^2 + \omega_i^2 \right] \left[ (Uk-\omega_r)^2 + \omega_i^2 + N^2 \sin^2 \theta \right]} \right\}, \label{stabcond2}
\end{equation}
where $U'^2_{max}={\rm Max}(U'^2)$.

We will now adopt a simple but systematic approach to obtain bounds for the expressions on the right-hand side of (\ref{stabcond2}). Defining $x=(Uk-\omega_r)^2$, we note that $x$ varies potentially between $x=0$ and $x \rightarrow \infty$. The extrema of any function of $x$ therefore occurs either at $x=0$, as $x \rightarrow \infty$, or at points where its derivative with respect to $x$ vanishes. The expression in curly brackets in (\ref{stabcond2}) can thus be written in terms of $x$ as 
\begin{equation}
\frac{(x-\omega_i^2 - N^2 \sin^2 \theta)^2 + 4\omega_i^2 x}{(x + \omega_i^2)(x+ \omega_i^2 + N^2 \sin^2 \theta)}. \label{defterm}
\end{equation}
The value of this expression at the point where its derivative with respect to $x$ vanishes is 
\begin{equation}
\frac{2 \omega_i^2 - N^2 \sin^2 \theta + \sqrt{(\omega_i^2+ N^2 \sin^2 \theta)(4 \omega_i^2 + N^2 \sin^2 \theta)}}{2 \omega_i^2 +\frac{5}{4} N^2 \sin^2 \theta + \sqrt{(\omega_i^2+ N^2 \sin^2 \theta)(4 \omega_i^2 + N^2 \sin^2 \theta)}}.
\label{lower}
\end{equation}
The value of (\ref{defterm}) at $x=0$ is $(\omega_i^2 + N^2 \sin^2 \theta)/\omega_i^2 = 1 + N^2 \sin^2 \theta/\omega_i^2 \ge 1$, and as $x \rightarrow +\infty$ it approaches $1$. On the other hand, (\ref{lower}) is never higher than $1$. 
This means that
\begin{equation}
{\rm Max} \left\{ \frac{\left[ (Uk-\omega_r)^2-\omega_i^2- N^2 \sin^2 \theta \right]^2 + 4 \omega_i^2 (Uk-\omega_r)^2}{\left[ (Uk-\omega_r)^2 + \omega_i^2 \right] \left[ (Uk-\omega_r)^2 + \omega_i^2 + N^2 \sin^2 \theta \right]} \right\} = 1 + \frac{N^2 \sin^2 \theta}{\omega_i^2}. \label{maxdef}
\end{equation}
Therefore, (\ref{stabcond2}) can be expressed as
\begin{equation}
4 \frac{N^2 \cos^2 \theta}{U'^2_{max}} > 1 + \frac{N^2 \sin^2 \theta}{\omega_i^2}. \label{stabcond3}
\end{equation}

Reversing the order of the above reasoning, if (\ref{stabcond3}) holds, then (\ref{stabcond}) is also satisfied, so the sum of the terms inside the outer round brackets in (\ref{condim}) is positive, and the instability growth rate $\omega_i$ must vanish.
Now, when $\omega_i=0$ and $\theta \neq 0$ (i.e., for non-vertical mean shear), the right-hand side of (\ref{stabcond3}) becomes infinite, making this condition impossible to satisfy. Therefore, it is not possible to obtain a stability condition in this situation, i.e., the flow may always be unstable for $\theta \neq 0$. For vertical shear, however ($\theta=0$), the second term on the right-hand side of (\ref{stabcond3}) drops, and the usual sufficient stability condition is obtained, namely: $Ri_{min} = N^2/U'^2_{max} > 1/4$. This shows that the general stability condition reduces to the Miles-Howard condition for vertical shear and is absent for non-vertical shear. The result is derived here for perturbations with 2D symmetry in planar shear flow using a method distinct from that of \cite{Fung_1986} and \cite{Deguchi_2021}. 

\subsection{Upper bound for the instability growth rate}

Returning to (\ref{condim}), if $\omega_i \neq 0$, then the sum of the terms inside the outer brackets must vanish, which implies
\begin{align}
&k^2 \int_{z_1}^{z_2} |G|^2 dz = \int_{z_1}^{z_2} \frac{k^2 U'^2}{|U k - \omega|^2} 
\left\{ \frac{1}{4} - \frac{N^2 \cos^2 \theta}{U'^2} \right. \nonumber \\ &\left. \times \frac{\left[(U k - \omega_r)^2 + \omega_i^2 \right] \left[ (U k -\omega_r)^2+ \omega_i^2+N^2 \sin^2 \theta \right]}{ \left[ (Uk-\omega_r)^2 - \omega_i^2 - N^2 \sin^2 \theta \right]^2 + 4 \omega_i^2 (Uk-\omega_r)^2} \right\} |G|^2 dz - \int_{z_1}^{z_2} |G'|^2 dz. \label{condgrowth}
\end{align}
On the other hand, (\ref{maxdef}) is equivalent to
\begin{equation}
{\rm Min} \left\{ \frac{\left[ (Uk-\omega_r)^2 + \omega_i^2 \right] \left[ (Uk-\omega_r)^2 + \omega_i^2 + N^2 \sin^2 \theta \right]}{\left[ (Uk-\omega_r)^2-\omega_i^2- N^2 \sin^2 \theta \right]^2 + 4 \omega_i^2 (Uk-\omega_r)^2} \right\} = \frac{\omega_i^2}{\omega_i^2 + N^2 \sin^2 \theta}, \label{mindef}
\end{equation}
so it follows from (\ref{condgrowth}) that
\begin{equation}
k^2 \int_{z_1}^{z_2} |G|^2 dz \le \int_{z_1}^{z_2} \frac{k^2 U'^2}{|Uk-\omega|^2} \left( \frac{1}{4} - \frac{N^2 \cos^2 \theta}{U'^2} \frac{\omega_i^2}{\omega_i^2 + N^2 \sin^2 \theta} \right) |G|^2 dz,
\end{equation}
because the integrals of $|G|^2$ and $|G'|^2$ are always positive. If we further note that $|U k - \omega|^2 \ge \omega_i^2$, then
\begin{equation}
\omega_i^2 \int_{z_1}^{z_2} |G|^2 dz \le \int_{z_1}^{z_2} U'^2 \left( \frac{1}{4} - \frac{N^2 \cos^2 \theta}{U'^2} \frac{\omega_i^2}{\omega_i^2 + N^2 \sin^2 \theta} \right) |G|^2 dz,
\end{equation}
which in turn implies, after division by the integral of $|G|^2$,
\begin{equation}
\omega_i^2 \le {\rm Max} \left( \frac{1}{4}U'^2 - \frac{\omega_i^2 N^2 \cos^2 \theta}{\omega_i^2 + N^2 \sin^2 \theta} \right).
\end{equation}
Since $N^2$ is assumed to be constant, this is equivalent to
\begin{equation}
\omega_i^2 \le \frac{1}{4}U'^2_{max} - \frac{\omega_i^2 N^2 \cos^2 \theta}{\omega_i^2 + N^2 \sin^2 \theta}.
\end{equation}
This inequality can be written as a second-order algebraic inequation for $\omega_i^2$, as follows:
\begin{equation}
\omega_i^4 + \left( N^2 - \frac{1}{4} U'^2_{max} \right) \omega_i^2 - \frac{1}{4} U'^2_{max} N^2 \sin^2 \theta \le 0. \label{eqomega}
\end{equation}
The solution to (\ref{eqomega}) can be shown to be
\begin{equation}
\omega_i^2 \le \frac{1}{2} \left\{ \frac{1}{4} U'^2_{max} - N^2 +\left[ \left( N^2- \frac{1}{4} U'^2_{max} \right)^2 + U'^2_{max} N^2 \sin^2 \theta \right]^{1/2} \right\}. \label{growth}
\end{equation}
This upper bound on the growth rate of the instability extends the formulation of \cite{Howard_1961} to flows with planar shear at an angle to the vertical. 
We now define a general Richardson number $Ri^*=N^2/U'^2$, allowing the mean shear to be inclined at any angle to the vertical, and $Ri^*_{min} = N^2/U'^2_{max}$. Then (\ref{growth}) can be expressed as
\begin{equation}
\frac{\omega_i^2}{U'^2_{max}} \le \frac{1}{2} \left\{ \frac{1}{4} - Ri^*_{min} + \left[ \left( Ri^*_{min} - \frac{1}{4} \right)^2 + Ri^*_{min} \sin^2 \theta \right]^{1/2} \right\}.
\label{growth2}
\end{equation}
Equation (\ref{growth2}) provides an upper bound for the normalized squared growth rate of the instability, $\omega_i^2/U'^2_{max}$. 
To our knowledge, the dependence of this upper bound on the shear angle, which is directly relevant to the numerical results of \cite{Candelier_etal_2011}, had not been derived previously.

\subsection{Extension of Kochar and Jain's semi-ellipse theorem}

\cite{Fung_1986} showed that the semi-ellipse theorem of  \cite{Kochar_Jain_1979} can be extended to flows with both horizontal and vertical shear, provided that the Richardson number used in the theorem is the minimum among those defined from the vertical and horizontal components of stratification and shear. In the present configuration, with purely vertical stratification, this minimum Richardson number is zero whenever horizontal shear is present, so that the semi-ellipse theorem reduces to the semi-circle theorem. The results obtained in the previous subsections, namely the absence of a stability condition and the independence of the instability growth rate from stratification for horizontal shear, are consistent with this behaviour. This suggests that the semi-ellipse theorem applies for vertical shear but reduces to the semi-circle theorem for horizontal shear, while its form for intermediate shear angles remains unclear. In the following, we extend the procedure of \cite{Kochar_Jain_1979} to planar shear at an arbitrary angle to the vertical in order to derive this dependence explicitly.

Following \cite{Miles_1961}, by introducing the change of variable $\hat{w}=(U k - \omega) F$, (\ref{tgeq}) becomes
\begin{equation}
\left[ (U k -\omega)^2 F' \right]' + k^2 (Uk-\omega)^2 \left[ \frac{N^2 \cos^2 \theta}{(Uk-\omega)^2 - N^2 \sin^2 \theta} - 1 \right] F = 0. \label{eqf}
\end{equation}
Multiplying this equation by $F^*$ and integrating it between $z_1$ and $z_2$ results in 
\begin{equation}
\int_{z_1}^{z_2} (Uk-\omega)^2 |F'|^2 dz - \int_{z_1}^{z_2} k^2 (Uk-\omega)^2 \left[ \frac{N^2 \cos^2 \theta}{(Uk-\omega)^2- N^2 \sin^2 \theta} - 1 \right] |F|^2 dz = 0, \label{intf1}   
\end{equation}
using integration by parts, together with the boundary conditions $F(z_1)=F(z_2)=0$. Equation (\ref{intf1}) can be rewritten as
\begin{equation}
\int_{z_1}^{z_2} (Uk-\omega)^2 R dz - \int_{z_1}^{z_2} \frac{N^2 k^2 \cos^2 \theta (Uk-\omega)^2}{(Uk-\omega)^2 - N^2 \sin^2 \theta} |F|^2 dz=0, \label{intf2}
\end{equation}
where $R=|F'|^2 + k^2 |F|^2$. The real and imaginary parts of (\ref{intf2}) are
\begin{align}
&\int_{z_1}^{z_2} \left[ (Uk-\omega_r)^2 
- \omega_i^2 \right] R dz 
- \int_{z_1}^{z_2} \left\{ P - Q \left[ (Uk-\omega_r)^2- \omega_i^2 \right] \right\} dz = 0, \label{eqfreal} \\
&-2 \omega_i \int_{z_1}^{z_2} (Uk-\omega_r) R dz
-2 \omega_i \int_{z_1}^{z_2} Q (Uk-\omega_r) dz = 0, \label{eqfimaginary}
\end{align}
where
\begin{align}
P &= \frac{N^2 k^2 \cos^2 \theta \left[ (Uk-\omega_r)^2 + \omega_i^2 \right]^2}{\left[ (Uk-\omega_r)^2 - \omega_i^2 - N^2 \sin^2 \theta \right]^2 + 4 \omega_i^2 (Uk-\omega_r)^2} |F|^2, \label{defp} \\
Q &= \frac{N^4 k^2 \sin^2 \theta \cos^2 \theta}{\left[ (Uk-\omega_r)^2 - \omega_i^2 - N^2 \sin^2 \theta \right]^2 + 4 \omega_i^2 (Uk-\omega_r)^2} |F|^2. \label{defq}
\end{align}
Equation (\ref{eqfimaginary}) can be expressed as
\begin{equation}
\omega_i \int_{z_1}^{z_2} (Uk-\omega_r) (R+Q) dz = 0.
\label{rqeq}
\end{equation}
Since $R \ge 0$ and $Q \ge 0$, if $\omega_i \neq 0$, $(Uk-\omega_r)$ must change sign between $z=z_1$ and $z=z_2$, so that the integral in (\ref{rqeq}) vanishes (a well-known result).
Adding (\ref{eqfreal}) and (\ref{eqfimaginary}) yields, after some manipulations,
\begin{equation}
\int_{z_1}^{z_2} U^2 k^2 (R+Q) dz = (\omega_r^2 + \omega_i^2) \int_{z_1}^{z_2} (R+Q) dz + \int_{z_1}^{z_2} P dz. \label{eqpqr2}
\end{equation}

Now, following \cite{Howard_1961}, since $(R+Q) \ge 0$, if $U_{min} \le U(z) \le U_{max}$ for $z_1 \le z \le z_2$, then
\begin{align}
&0 \ge \int_{z_1}^{z_2} k^2 (U-U_{min})(U-U_{max}) (R+Q) dz = \int_{z_1}^{z_2} U^2 k^2 (R+Q)dz \nonumber \\ 
&- (U_{min}+U_{max}) k^2 \int_{z_1}^{z_2} U (R+Q) dz + U_{min} U_{max} k^2 \int_{z_1}^{z_2} (R+Q) dz.
\label{semic}
\end{align}
Using (\ref{rqeq}) and (\ref{eqpqr2}) in (\ref{semic}), the latter equation becomes
\begin{equation}
\left[ \omega_r^2 + \omega_i^2 -(U_{min}+U_{max})k \omega_r +U_{min} U_{max} k^2 \right] \int_{z_1}^{z_2} (R + Q) \le -\int_{z_1}^{z_2} P dz.
\label{semic3}
\end{equation}
Noting that the integral involving $P$ in this equation is non-negative yields the semi-circle theorem of \cite{Howard_1961}. However, a more restrictive condition can be obtained following the approach of \cite{Kochar_Jain_1979}. 

The functions $F$ and $G$ are related through $G=(U k - \omega)^{1/2} F$, from which it follows that 
\begin{equation}
|G'|^2 \ge |Uk-\omega| |F'|^2 + \frac{1}{4} \frac{U'^2 k^2}{|Uk-\omega|} |F|^2 - |U'|k |F| |F'|.
\label{inegl}
\end{equation}
This is analogous to the equation in section 3 of \cite{Kochar_Jain_1979}. Integrating (\ref{inegl}) with respect to $z$ between $z_1$ and $z_2$ gives
\begin{equation}
\int_{z_1}^{z_2} |G'|^2 dz \ge \int_{z_1}^{z_2} |U k - \omega| |F'|^2 dz + \frac{1}{4} k^2 \int_{z_1}^{z_2} \frac{U'^2}{|Uk-\omega|} |F|^2 dz - k \int_{z_1}^{z_2} |U'| |F| |F'| dz.
\label{intinegl}
\end{equation}

Considering the unstable case in which $\omega_i \neq 0$, (\ref{condgrowth}) can be expressed as
\begin{align}
&\int_{z_1}^{z_2} |G'|^2 dz = \frac{1}{4} k^2 \int_{z_1}^{z_2} \frac{U'^2}{|U k - \omega|} 
\left\{ 1 -4 \frac{N^2 \cos^2 \theta}{U'^2} \right. \nonumber \\ 
&\left. \times \frac{\left[(U k - \omega_r)^2 + \omega_i^2 \right] \left[ (U k -\omega_r)^2+ \omega_i^2+N^2 \sin^2 \theta \right]}{ \left[ (Uk-\omega_r)^2 - \omega_i^2 - N^2 \sin^2 \theta \right]^2 + 4 \omega_i^2 (Uk-\omega_r)^2} \right\} |F|^2 dz - k^2 \int_{z_1}^{z_2} |Uk-\omega| |F|^2 dz.
\label{condgrowth2}
\end{align}
Equations (\ref{intinegl}) and (\ref{condgrowth2}) then imply that
\begin{align}
&\frac{1}{4} k^2 \int_{z_1}^{z_2} \frac{U'^2}{|U k - \omega|} 
\left\{ 1 -4 \frac{N^2 \cos^2 \theta}{U'^2} \frac{\left[(U k - \omega_r)^2 + \omega_i^2 \right] \left[ (U k -\omega_r)^2+ \omega_i^2+N^2 \sin^2 \theta \right]}{ \left[ (Uk-\omega_r)^2 - \omega_i^2 - N^2 \sin^2 \theta \right]^2 + 4 \omega_i^2 (Uk-\omega_r)^2} \right\} \nonumber \\ 
&\times |F|^2 dz - k^2 \int_{z_1}^{z_2} |Uk-\omega| |F|^2 dz \ge \int_{z_1}^{z_2} |U k - \omega| |F'|^2 dz \nonumber \\
&+ \frac{1}{4} k^2 \int_{z_1}^{z_2} \frac{U'^2}{|Uk-\omega|} |F|^2 dz - k \int_{z_1}^{z_2} |U'| |F| |F'| dz.
\label{interm1}
\end{align}
On the other hand, (\ref{mindef}) allows (\ref{interm1}) to be written as
\begin{align}
&\frac{1}{4} k^2 \int_{z_1}^{z_2} \frac{U'^2}{|U k - \omega|} 
\left[ 1 - 4 \frac{N^2 \cos^2\theta \omega_i^2}{U'^2 ( \omega_i^2 + N^2 \sin^2 \theta )} \right]  |F|^2 dz \nonumber \\
&\ge \int_{z_1}^{z_2} |Uk-\omega| R dz + \frac{1}{4} k^2 \int_{z_1}^{z_2} \frac{U'^2}{|Uk-\omega|} |F|^2 dz - k \int_{z_1}^{z_2} |U'| |F| |F'| dz.
\label{interm2}
\end{align}
The Schwarz inequality may be used to estimate the last term on the right of (\ref{interm2}) as
\begin{align}
\int_{z_1}^{z_2} |U'| |F| |F'| &\le \left( \int_{z_1}^{z_2} \frac{U'^2}{|Uk-\omega|} |F|^2 dz \right)^{1/2} \left( \int_{z_1}^{z_2} |Uk-\omega| |F'|^2 dz \right)^{1/2} \nonumber \\
&\le  \left( \int_{z_1}^{z_2} \frac{U'^2}{|Uk-\omega|} |F|^2 dz \right)^{1/2} \left( \int_{z_1}^{z_2} |Uk-\omega| R dz \right)^{1/2},
\label{schwartz}
\end{align}
which may be used to deduce from (\ref{interm2}) that
\begin{align}
&\left[  1 - 4 \frac{N^2 \cos^2 \theta \omega_i^2}{U'^2_{max} ( \omega_i^2 + N^2 \sin^2 \theta)} \right] \frac{1}{4} k^2 \int_{z_1}^{z_2} \frac{U'^2}{|U k - \omega|}  |F|^2 dz \nonumber \\
&\ge \int_{z_1}^{z_2} |Uk-\omega| R dz + \frac{1}{4} k^2 \int_{z_1}^{z_2} \frac{U'^2}{|Uk-\omega|} |F|^2 dz - k \left( \int_{z_1}^{z_2} \frac{U'^2}{|Uk-\omega|} |F|^2 dz \right)^{1/2} \nonumber \\
&\times\left( \int_{z_1}^{z_2} |Uk-\omega| R dz \right)^{1/2}.
\label{interm3}
\end{align}
Defining
\begin{equation}
B^2 = \frac{1}{4} k^2 \int_{z_1}^{z_2} \frac{U'^2}{|U k - \omega|}  |F|^2 dz, \quad E^2 = \int_{z_1}^{z_2} |Uk-\omega| R dz,
\label{defbe}
\end{equation}
(\ref{interm3}) can be expressed as a 2nd order algebraic inequality, as follows:
\begin{equation}
\left( \frac{E}{B} \right)^2 - 2 \frac{E}{B} + 4 Ri^*_{min} \frac{\omega_i^2 \cos^2 \theta}{\omega_i^2 + N^2 \sin^2 \theta} \le 0.
\label{2ndorder}
\end{equation}
The solution of (\ref{2ndorder}) is 
\begin{equation}
1-\sqrt{1 - 4 Ri^*_{min} \frac{\omega_i^2 \cos^2 \theta}{\omega_i^2+ N^2 \sin^2 \theta}} \le \frac{E}{B} \le 1+\sqrt{1 - 4 Ri^*_{min} \frac{\omega_i^2 \cos^2 \theta}{\omega_i^2+ N^2 \sin^2 \theta}},
\label{solution}
\end{equation}
with the right-hand relation being equivalent to
\begin{equation}
\int_{z_1}^{z_2} |Uk-\omega| R dz \le \left[ 1+\sqrt{1 - 4 Ri^*_{min} \frac{\omega_i^2 \cos^2 \theta}{\omega_i^2+ N^2 \sin^2 \theta}} \right]^2 \frac{1}{4} k^2 \int_{z_1}^{z_2} \frac{U'^2}{|U k - \omega|}  |F|^2 dz.
\label{interm4}
\end{equation}
Taking into account that $|Uk-\omega| \ge \omega_i$, (\ref{interm4}) implies
\begin{align}
&\int_{z_1}^{z_2} R  dz \le \frac{1}{4} \frac{k^2}{\omega_i^2} \left[ 1+\sqrt{1 - 4 Ri^*_{min} \frac{\omega_i^2 \cos^2 \theta}{\omega_i^2+ N^2 \sin^2 \theta}} \right]^2 \int_{z_1}^{z_2} U'^2 |F|^2 dz \nonumber \\
&\le \frac{1}{4} \frac{k^2 U'^2_{max}}{\omega_i^2} \left[ 1+\sqrt{1 - 4 Ri^*_{min} \frac{\omega_i^2 \cos^2 \theta}{\omega_i^2+ N^2 \sin^2 \theta}} \right]^2 \int_{z_1}^{z_2} |F|^2 dz.
\label{interm5}
\end{align}

To proceed from (\ref{semic3}), a lower bound for the integral involving $P$ is required.
Using the method outlined in section 2.2, note that $P$ attains its minimum at $x=0$ and from (\ref{defp}) it follows that
\begin{equation} 
P \ge \frac{N^2 k^2 \cos^2 \theta \omega_i^4}{(\omega_i^2 + N^2 \sin^2 \theta)^2} |F|^2,
\label{minp}
\end{equation}
which implies
\begin{equation}
\int_{z_1}^{z_2} P dz \ge \frac{N^2 k^2 \cos^2 \theta \omega_i^4}{(\omega_i^2 + N^2 \sin^2 \theta)^2} \int_{z_1}^{z_2} |F|^2 dz.
\label{minintp}
\end{equation}
An upper bound for the integral of $Q$ is also required. From its definition in (\ref{defq}), and using the same method, it follows that
\begin{align}
&Q \le \frac{N^4 k^2 \sin^2 \theta \cos^2 \theta}{4 \omega_i^2 N^2 \sin^2 \theta} |F|^2 \quad {\rm if} \quad N^2 \sin^2 \theta \ge \omega_i^2, \nonumber \\
&Q \le \frac{N^4 k^2 \sin^2 \theta \cos^2 \theta}{(\omega_i^2 + N^2 \sin^2 \theta)^2} |F|^2 \quad {\rm if} \quad N^2 \sin^2 \theta \le \omega_i^2.
\label{maxq}
\end{align}
Both conditions are satisfied if
\begin{equation}
Q \le \frac{N^4 k^2 \sin^2 \theta \cos^2 \theta}{\omega_i^4} |F|^2,
\label{maxq2}
\end{equation}
so that
\begin{equation}
\int_{z_1}^{z_2} Q dz \le \frac{N^4 k^2 \sin^2 \theta \cos^2 \theta}{\omega_i^4} \int_{z_1}^{z_2} |F| dz.
\label{maxintq}
\end{equation}
Adding (\ref{interm5}) and (\ref{maxintq}) and rearranging yields
\begin{align}
&\int_{z_1}^{z_2} |F|^2 dz \ge \frac{\omega_i^2}{N^2 k^2} \left\{ \frac{1}{4 Ri^*_{min}} \left[ 1+\sqrt{1 - 4 Ri^*_{min} \frac{\omega_i^2 \cos^2 \theta}{\omega_i^2 + N^2 \sin^2 \theta}} \right]^2 + \frac{N^2 \sin^2 \theta \cos^2 \theta}{\omega_i^2} \right\}^{-1} \nonumber \\ 
&\times \int_{z_1}^{z_2} (R + Q) dz.
\label{invintrq}
\end{align} 
Substituting (\ref{invintrq}) into (\ref{minintp}) gives
\begin{align}
&\int_{z_1}^{z_2} P dz \ge \frac{\omega_i^6 \cos^2 \theta}{(\omega_i^2 + N^2 \sin^2 \theta)^2} \nonumber \\
&\times \left\{ \frac{1}{4 Ri^*_{min}} \left[ 1+\sqrt{1 - 4 Ri^*_{min} \frac{\omega_i^2 \cos^2 \theta}{\omega_i^2 + N^2 \sin^2 \theta}} \right]^2 + \frac{N^2 \sin^2 \theta \cos^2 \theta}{\omega_i^2} \right\}^{-1} \int_{z_1}^{z_2} (R + Q) dz.
\label{boundp}
\end{align}

Finally, substituting (\ref{boundp}) into (\ref{semic3}) for the term involving $P$, dividing by the integral of $(R+Q)$ (which then appears in every term) and rearranging yields
\begin{align}
&\left( \omega_r - k \frac{U_{max}+U_{min}}{2} \right)^2 \nonumber \\
&+\omega_i^2 \left( 1+ \frac{4 Ri^*_{min} \omega_i^4 \cos^2 \theta}{ \left( \omega_i^2  + N^2 \sin^2 \theta \right)^2 \left\{ \left[ 1+\sqrt{1 - 4 Ri^*_{min} \frac{\omega_i^2 \cos^2 \theta}{\omega_i^2 + N^2 \sin^2 \theta}} \right]^2 + \frac{4 Ri^*_{min} N^2 \sin^2 \theta \cos^2 \theta}{\omega_i^2} \right\}} \right) \nonumber \\
&\le k^2 \left( \frac{U_{max}-U_{min}}{2} \right)^2. 
\label{semiell2}
\end{align}
This new result defines bounds for $\omega_r + i \omega_i$ in the complex semi-plane that are, in general, no longer strictly circular or elliptical, due to the additional dependences on $\omega_i$ in the second line of (\ref{semiell2}).

\section{Discussion of results}

The results derived above provide an explicit dependence of the integral constraints placed on the growth rate and complex angular frequency of the addressed instability on the shear angle, being directly comparable with the numerical results of \cite{Candelier_etal_2011}, which use the same assumptions. Equation (\ref{stabcond3}) expresses the usual Miles-Howard stability condition for flow with vertical shear, and the absence of a stability condition for shear with any horizontal component. An analogous result applicable to arbitrary parallel non-planar shear flows was sketched in an incomplete form by \cite{Fung_1986}, and presented in full by \cite{Deguchi_2021}. Subject to its different assumptions, the present result is equivalent to these previous ones, but the method used to obtain it was different. 

As far as we know, all other results derived above have not been obtained previously in any form.
The upper bound of the growth rate of the linear instability derived by \cite{Howard_1961} is extended by (\ref{growth2}) to shear with an orientation between vertical and horizontal.  Equation (\ref{semiell2}), provides a dependence of the form of a theorem analogous to the Miles-Howard semi-circle theorem or the Kochar-Jain semi-ellipse theorem on the angle of the shear with the vertical, which is only possible for planar shear, and in that sense goes beyond the extension by \cite{Fung_1986} of the semi-ellipse theorem of \cite{Kochar_Jain_1979} to a generic parallel shear flow. We will discuss next the flow behaviour associated with these two original results. 

\subsection{Upper bound for the instability growth rate}

The extended formula obtained for the upper bound of the instability growth rate, (\ref{growth}), has interesting properties despite its simple form. It reduces, as expected, to Howard's corresponding result for $\theta=0$ (vertical shear), but for $\theta=\pi/2$ (horizontal shear), it becomes
\begin{equation}
\omega_i^2 \le \frac{1}{4} U'^2_{max}. \label{growth3}
\end{equation}
This means that the upper bound of the growth rate no longer depends of $N$, which is consistent with the absence of $N$ in (\ref{tgeq}) in this flow configuration. Equation (\ref{growth3}) is equivalent to equation (4.10) of \cite{Hoiland_1953} (see also \cite{Drazin_Howard_1966}), valid for a non-stratified fluid, and is also equal to the limit of (\ref{growth}) for $N=0$.
In dimensionless form, this corresponds to the limit of (\ref{growth2}) for $\theta=\pi/2$ or $Ri^*_{min}=0$. At the opposite limit, $Ri^*_{min} \rightarrow +\infty$, it can be shown that (\ref{growth}) reduces to
\begin{equation}
\omega_i^2 \le \frac{1}{4} U'^2_{max} \sin^2 \theta. \label{growth4}
\end{equation}

The best term of comparison for (\ref{growth}) is figure 4 of \cite{Candelier_etal_2011}, which shows the instability growth rate for 2D perturbations (the most unstable in this context) in a stratified flow with planar shear at an arbitrary angle to the vertical, as a function of that angle $\theta$ and the Froude number associated with the maximum shear, $F$. For the particular velocity profile adopted by \cite{Candelier_etal_2011} (a 2D Bickley jet), $F$ is related to the minimum Richardson number defined above, as $Ri^*_{min}=27/(16 F^2)$, where $F=U_0/(LN)$, and $L$ and $U_0$ are characteristic length and velocity scales of the flow, respectively. It should also be noted that all quantities in \cite{Candelier_etal_2011} are normalized by $L$ and $U_0$, and the growth rate shown in their figure 4 is $\omega_i L/U_0 = [4/(3 \sqrt{3})] \omega_i/|U'_{max}|$. Using these relations, if the upper bound for the growth rate of the instability is denoted by $\omega_{imax}$, (\ref{growth2}) may be expressed as
\begin{equation}
\frac{\omega_{imax}^2 L^2}{U_0^2} = \frac{16}{27} \left\{ \frac{1}{2} \left( \frac{1}{4} - \frac{27}{16 F^2} \right) + \frac{1}{2} \left[ \left( \frac{1}{4} - \frac{27}{16 F^2} \right)^2 + \frac{27}{16 F^2} \sin^2 \theta \right]^{1/2} \right\}. \label{growth5}
\end{equation}
\begin{figure}[h!]
\centerline{\includegraphics[width=80mm]{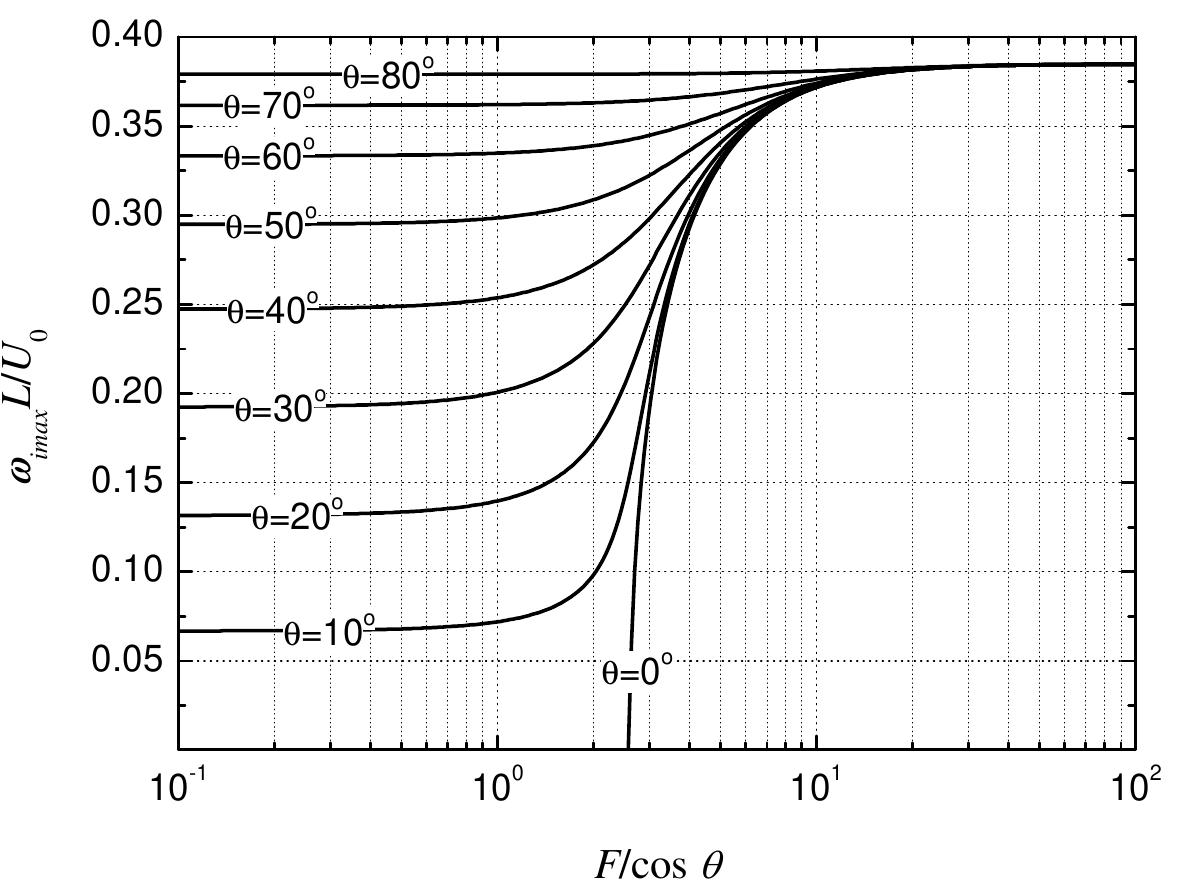}}
\caption{Normalized upper bound $\omega_{imax} L/U_0$ of the growth rate of the linear instability of perturbations with 2D symmetry in planar stratified flow with shear at an angle $\theta$ to the vertical, given by (\ref{growth5}) as a function of $F/\cos \theta$, for different values of $\theta$ (see line labels for details).}
\label{fig1}
\end{figure}
This relation is shown in figure \ref{fig1} in the same form as in figure 4 of \cite{Candelier_etal_2011}. For $\theta=0$, the upper bound of $\omega_i L/U_0$ is zero for $F<3\sqrt{3}/2$, which corresponds to $Ri^*_{min}>1/4$, implying flow stability. When $F\gg 1$, which corresponds to $Ri^*_{min} \ll 1$, $\omega_{imax} L/U_0$ asymptotically approaches $\approx0.38$ for all values of $\theta$, which corresponds to $\omega_{imax}^2/U'^2_{max}=1/4$. On the other hand, as $\theta$ increases, $\omega_{imax} L/U_0$ at low $F$ (which corresponds to high $Ri^*_{min}$) increases from zero for $\theta=0$ to $\approx 0.38$ in the limit $\theta=\pi/2$. This behaviour is qualitatively similar to that shown in figure 4 of \cite{Candelier_etal_2011}. The difference is that the actual growth rate $\omega_i$ is considerably smaller, by a factor of $\approx 2.4$, than its upper bound $\omega_{imax}$, shown in figure \ref{fig1}. Additionally, in figure 4 of \cite{Candelier_etal_2011}, $\omega_i$ has a minimum near $F=3 \sqrt{3}/2$ before increasing to its maximum at larger $F$, whereas in figure \ref{fig1}, $\omega_{imax}$ increases monotonically with $F$.

 From (\ref{growth5}), we may deduce, (consistently with (\ref{growth3})--(\ref{growth4})), that
\begin{equation}
\frac{\omega_{imax}(F = 0)}{\omega_{imax}(F \rightarrow +\infty)} = \frac{\omega_{imax}(Ri^*_{min} \rightarrow +\infty)}{\omega_{imax} (Ri^*_{min} = 0)} = \sin \theta.
\label{asympt}
\end{equation}
Interestingly, this result is also exactly satisfied by the growth rate $\omega_i$ shown in figure 4 of \cite{Candelier_etal_2011}, and is consistent with their low $F$ asymptotic theory. 

It is worth  noting that if, in addition to $l=0$, we also assumed that $v=0$ in (\ref{eqmomu})-(\ref{eqmass}), i.e., for fully 2D flow, (\ref{tgeq}) would differ from its traditional version valid for vertical shear only by a factor of $\cos^2 \theta$ multiplying the term proportional to $N^2$. In that case, the original results of Miles and Howard would remain applicable, provided that $N^2$ was replaced by $N^2 \cos^2 \theta$, or that $Ri^*_{min}$ was replaced by $Ri^*_{min} \cos^2 \theta$ (cf.~\citealp{Jacobitz_Sarkar_1998}). This would lead to an upper bound on the squared growth rate of the instability given by
\begin{equation}
\frac{\omega_i^2}{U'^2_{max}} \le  \frac{1}{4} - Ri^*_{min} \cos^2 \theta,
\label{wrong}
\end{equation}
corresponding to a stability condition $Ri^*_{min} >1/(4 \cos^2 \theta)$, which, when expressed in terms of $F$, is clearly incompatible with figure 4 of \cite{Candelier_etal_2011}.

\subsection{Extension of the semi-ellipse theorem}

The extension of the semi-ellipse theorem of Kochar and Jain expressed by (\ref{semiell2}) reduces to Howard's semi-circle theorem when either $Ri^*_{min}=0$ (no stratification) or $\theta=\pi/2$ (horizontal shear), and to Kochar and Jain's semi-ellipse theorem for $\theta=0$ (vertical shear), as anticipated. In intermediate situations, it exhibits a richer behaviour, which is illustrated and discussed next.

Equation (\ref{semiell2}) is expressed in terms of $\omega_r$ and $\omega_i$, rather than the more usual phase speed variables $c_r$ and $c_i=\omega_i/k$, which reduces its explicit dependence on $k$. 
With this choice, a natural normalization of $\omega$ is obtained using $N$, leading to the following dimensionless form of (\ref{semiell2}):
\begin{align}
&\left(\frac{\omega_r}{N} -\frac{k}{N} \frac{U_{max}+U_{min}}{2} \right)^2 \nonumber \\
&+\left( \frac{\omega_i}{N} \right)^2 \left( 1+ \frac{4 Ri^*_{min} \left( \frac{\omega_i}{N} \right)^4 \cos^2 \theta}{ \left[ \left(\frac{\omega_i}{N} \right)^2  + \sin^2 \theta \right]^2 \left\{ \left[ 1+\sqrt{1 - 4 Ri^*_{min} \frac{\left( \omega_i/N \right)^2 \cos^2 \theta}{\left( \omega_i/N \right)^2 + \sin^2 \theta}} \right]^2 + \frac{4 Ri^*_{min} \sin^2 \theta \cos^2 \theta}{\left( \omega_i/N \right)^2} \right\}} \right) \nonumber \\
&\le \left( \frac{k}{N} \frac{U_{max}-U_{min}}{2} \right)^2. 
\label{semiell3}
\end{align}
The values of $(\omega_r/N,\omega_i/N)$ in the complex plane corresponding to the bounds in (\ref{semiell3}) are obtained by replacing the inequality by equality signs. These values depend on $\theta$, $Ri^*_{min}$, $k U_{min}/N$ and $k U_{max}/N$. For illustration, we focus on the case $k U_{min}/N =0$, which corresponds to a localized shear flow, such as the Bickley jet considered by \cite{Candelier_etal_2011}.

An important difference between (\ref{semiell3}) and the usual semi-ellipse ($\theta=0$) and semi-circle ($\theta=\pi/2$) theorems is the more complicated dependence on $\omega_i$. In particular, the presence of $\omega_i$ inside the square-root in the second line introduces an additional restriction on this quantity in addition to the condition imposed by (\ref{growth}), requiring the argument of the square-root to be non-negative (in the semi-ellipse case this condition reduces to $Ri^*_{min} \le 1/4$). This leads in general to the condition
\begin{equation} 
\left( \frac{\omega_i}{N} \right)^2 \le \frac{\sin^2 \theta}{4 Ri^*_{min} \cos^2 \theta -1} \quad {\rm if} \quad Ri^*_{min} \ge \frac{1}{4 \cos^2 \theta},
\label{condpos}
\end{equation}
with no restriction necessary otherwise. This means, more specifically, that if $Ri^*_{min} < 1/4$, the argument of the square-root is always non-negative for any $\theta$.

Equation (\ref{condpos}) highlights a closer relationship between (\ref{growth}) and (\ref{semiell2}) than in the traditional semi-ellipse or semi-circle formulations. It is therefore convenient to normalize (\ref{growth}) in a similar way, using $N$ (instead of $U'_{max}$, as in (\ref{growth2})), which gives
\begin{equation}
\left( \frac{\omega_i}{N} \right)^2 \le \frac{1}{2} \left\{ \frac{1}{4 Ri^*_{min}} - 1 + \left[ \left( 1 - \frac{1}{4 Ri^*_{min}} \right)^2 + \frac{\sin^2 \theta}{Ri^*_{min}} \right]^{1/2} \right\}.
\label{growth6}
\end{equation}
Note that, compared with $\omega_i^2/U'^2_{max}$ in (\ref{growth2}), $\omega_i^2/N^2$ in (\ref{growth6}) is not bounded as $Ri^*_{min} \rightarrow 0$. More importantly for the present discussion, (\ref{growth6}) imposes a more restrictive condition than (\ref{condpos}), so that if (\ref{growth6}) is satisfied, (\ref{condpos}) is automatically satisfied. This interdependence does not arise in the traditional semi-ellipse or semi-circle theorems (corresponding to $\theta=0$ and $\theta=\pi/2$ in (\ref{semiell3}), respectively). 

\begin{figure}[h!]
\centerline{
\includegraphics[width=36mm]{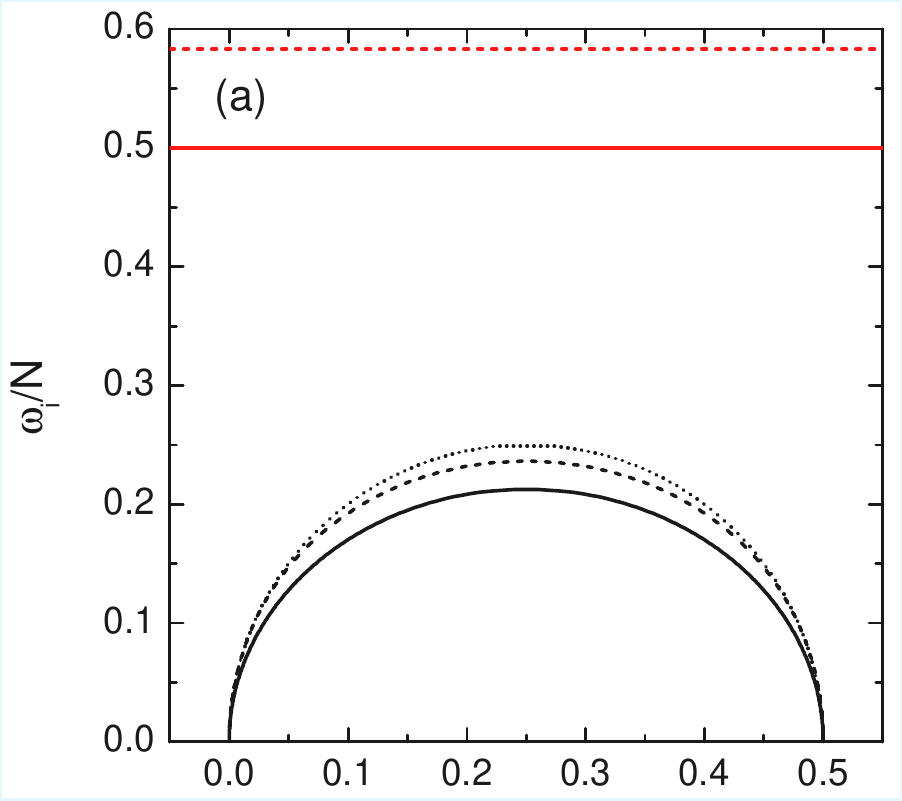}
\includegraphics[width=32mm]{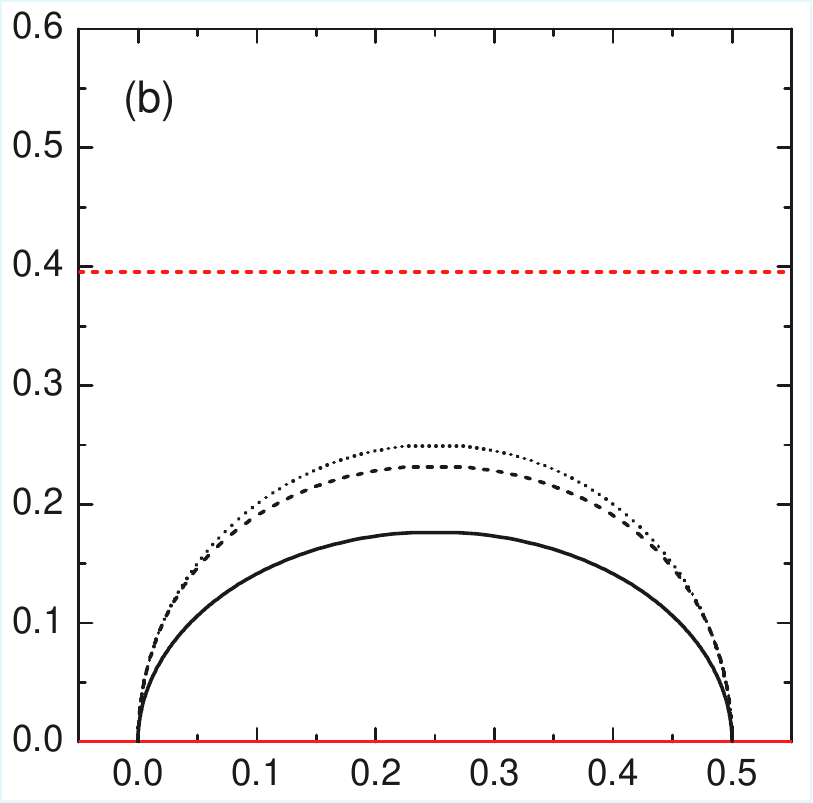}
\includegraphics[width=32mm]{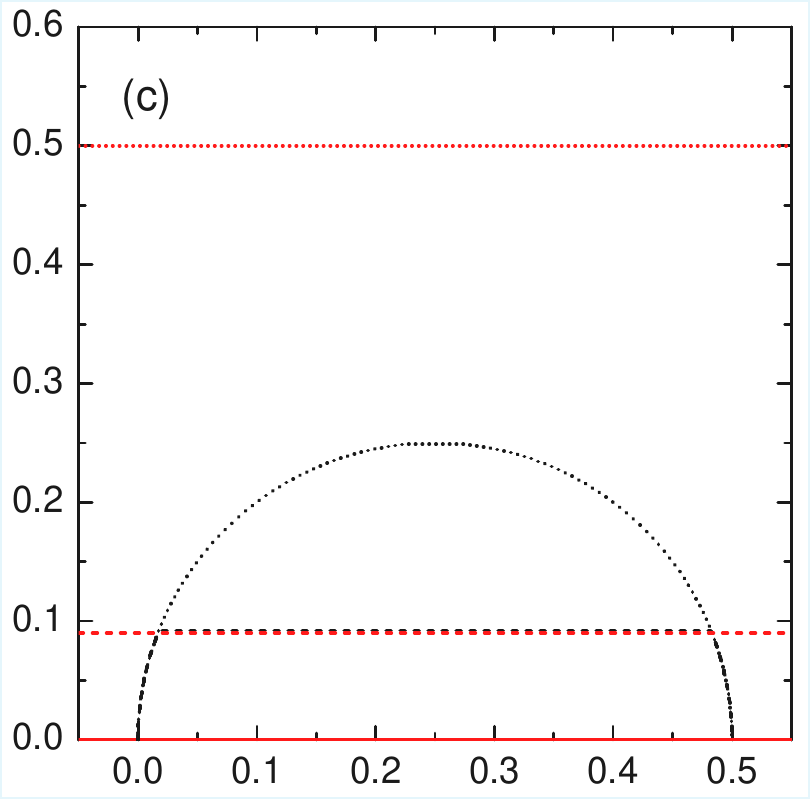}
\includegraphics[width=32mm]{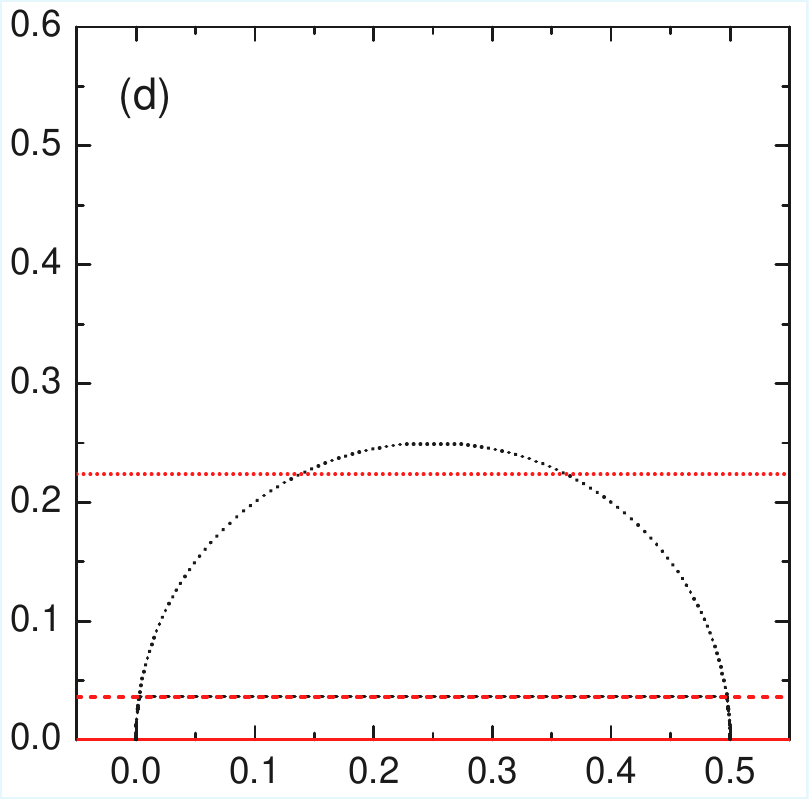}}
\centerline{
\includegraphics[width=36mm]{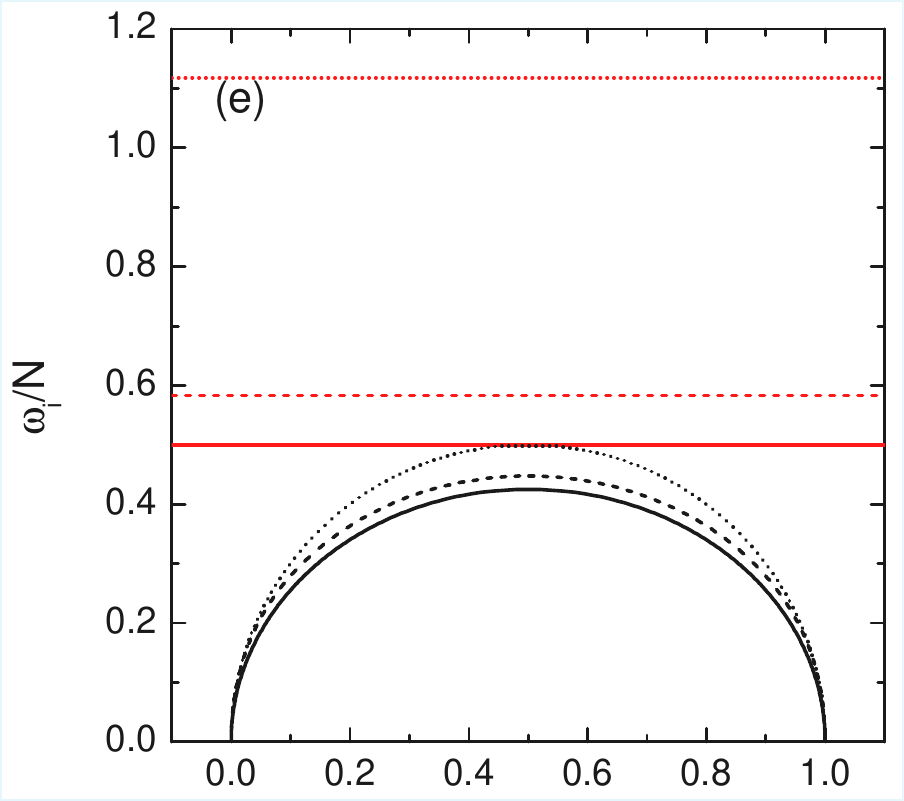}
\includegraphics[width=32mm]{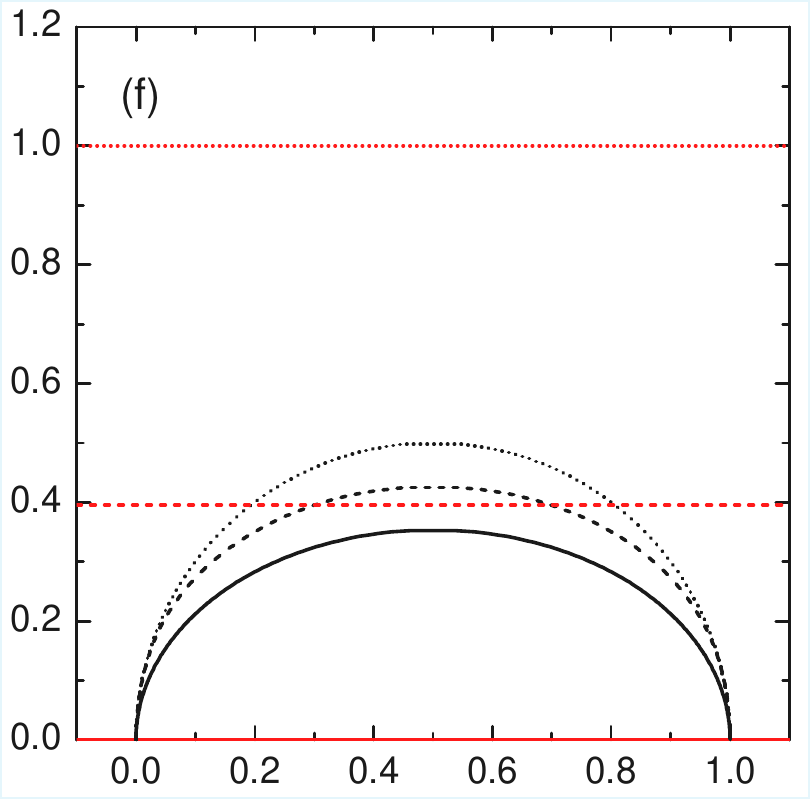}
\includegraphics[width=32mm]{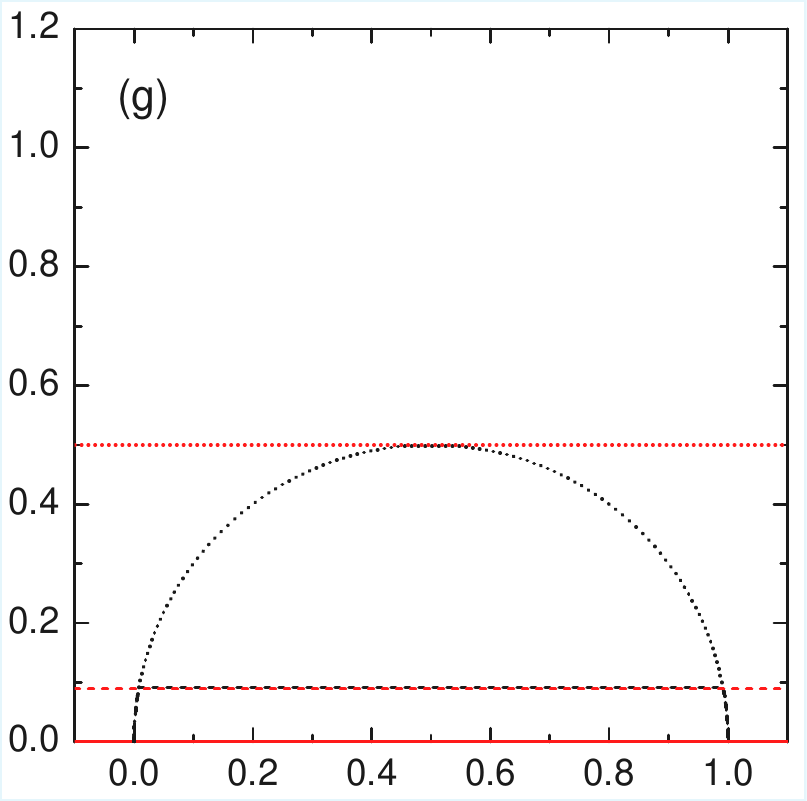}
\includegraphics[width=32mm]{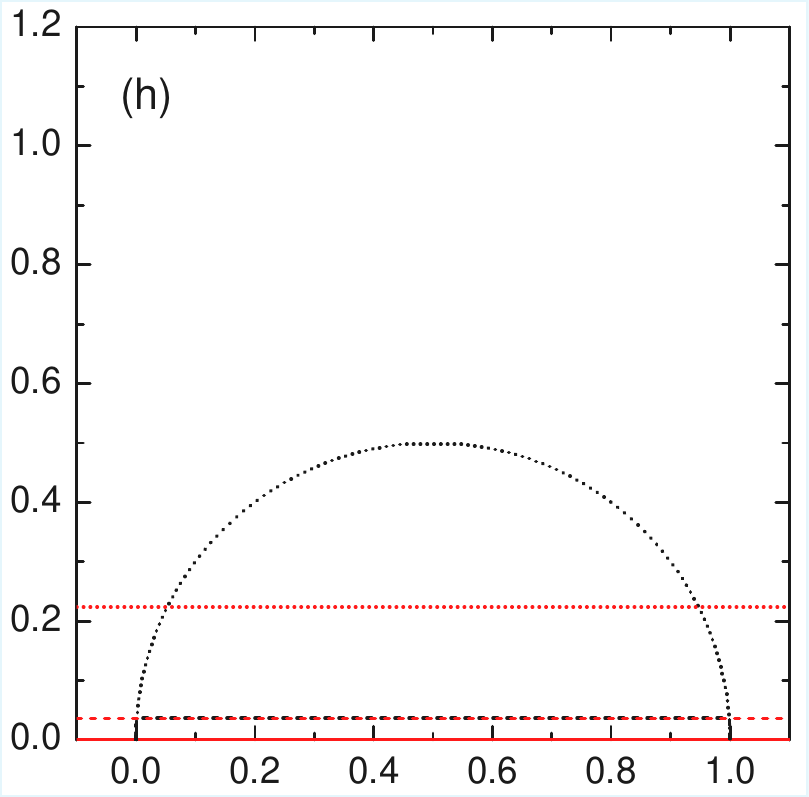}}
\centerline{
\includegraphics[width=36mm]{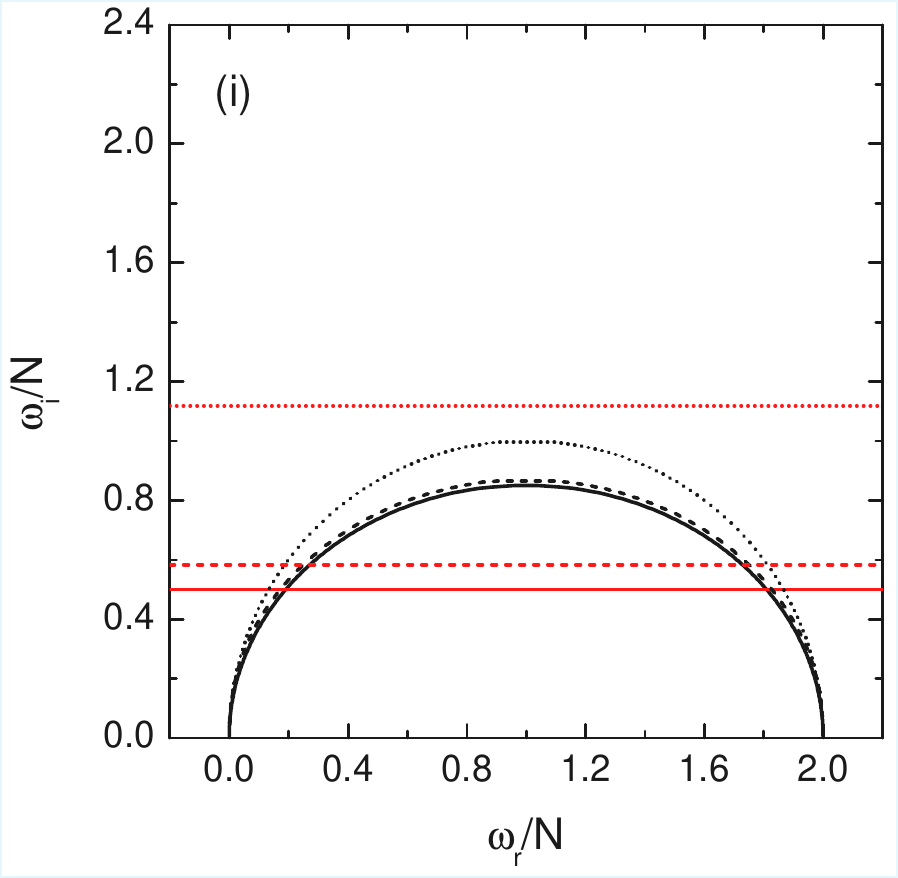}
\includegraphics[width=32mm]{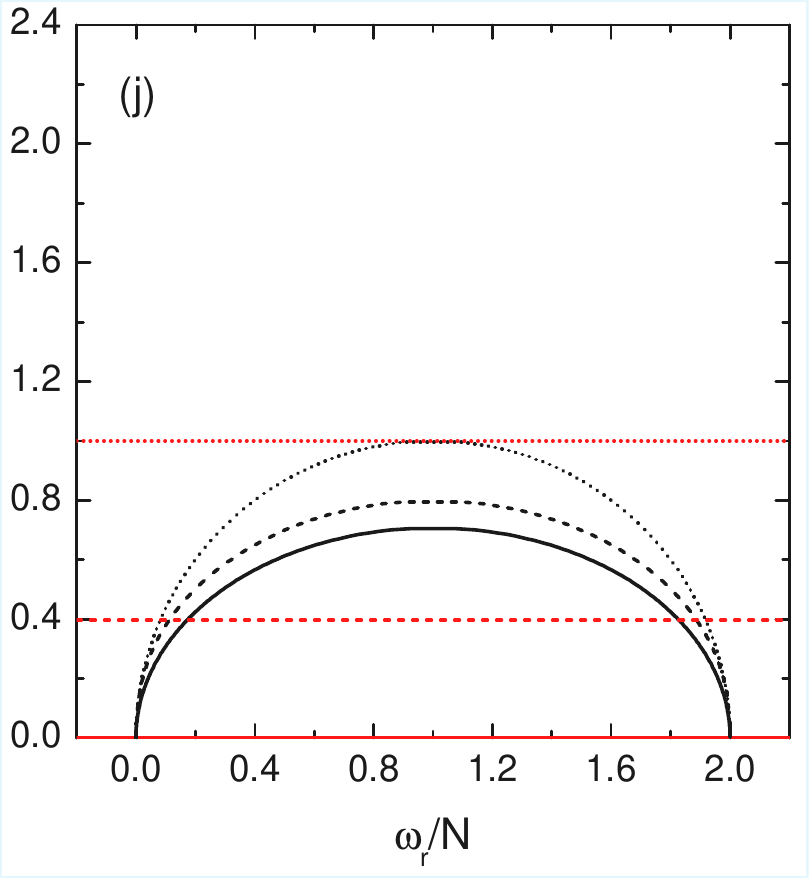}
\includegraphics[width=32mm]{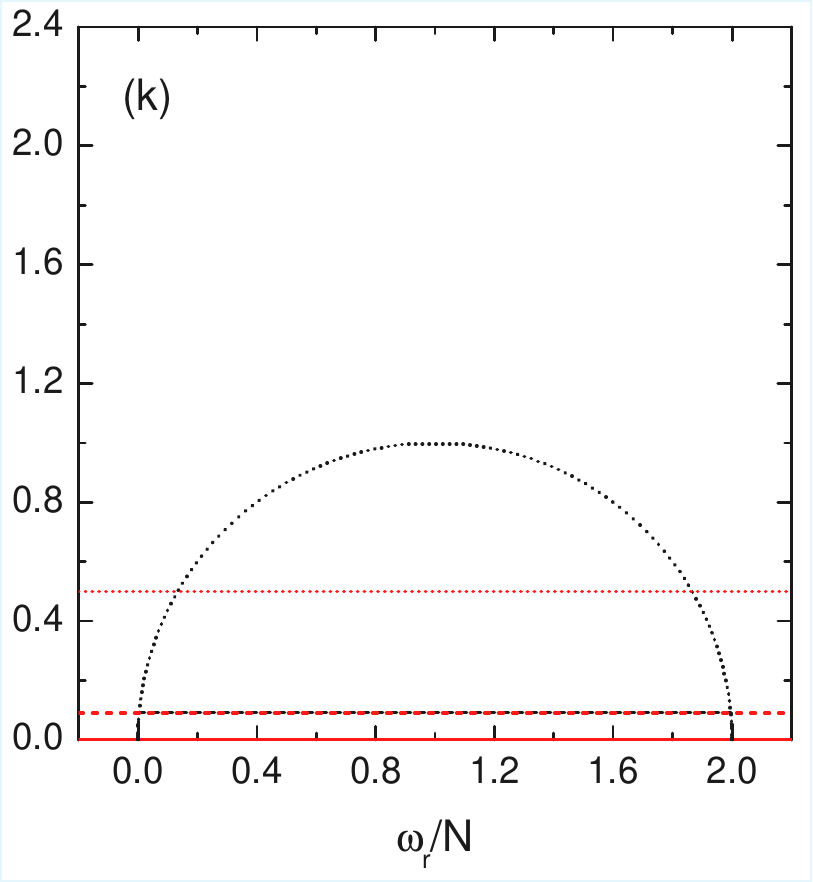}
\includegraphics[width=32mm]{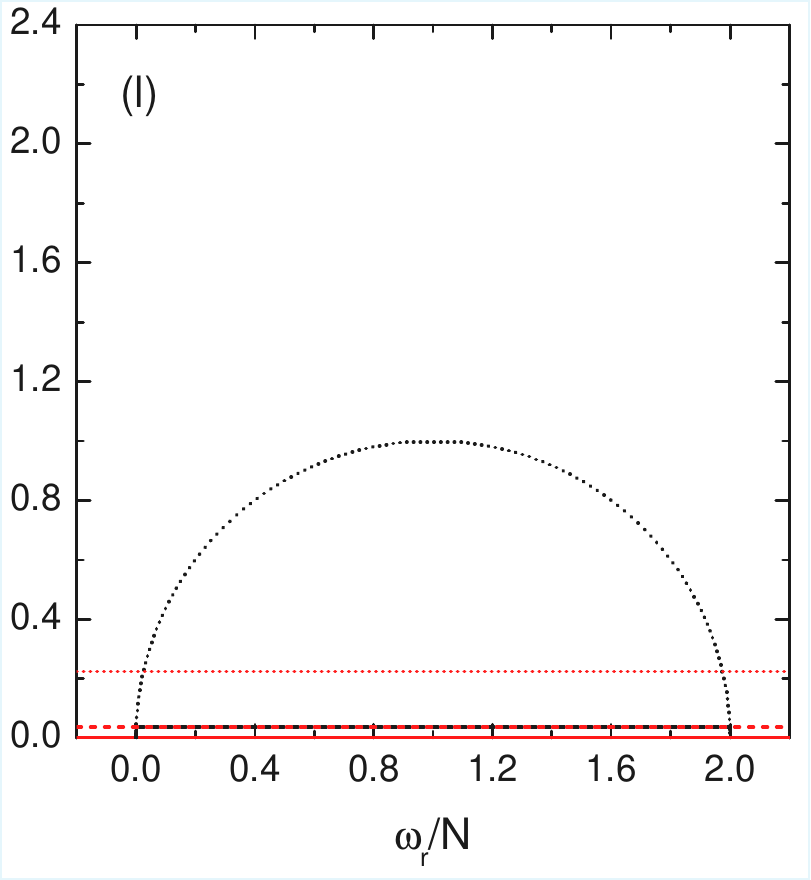}}
\caption{Bounds imposed on $(\omega_r/N,\omega_i/N)$ by (\ref{growth6}) (red lines) and the extension of the semi-ellipse theorem (\ref{semiell3}) (black lines) for $k U_{min}/N=0$. Solid lines $\theta=0$, dashed lines: $\theta=\pi/20$, dotted lines: $\theta=\pi/2$. (a) $Ri^*_{min}=0.2$, $U_{max} k/N=0.5$; (b) $Ri^*_{min}=0.25$, $U_{max} k/N=0.5$; (c) $Ri^*_{min}=1$, $U_{max} k/N=0.5$; (d) $Ri^*_{min}=5$, $U_{max} k/N=0.5$; (e) $Ri^*_{min}=0.2$, $U_{max} k/N=1$; (f) $Ri^*_{min}=0.25$, $U_{max} k/N=1$; (g) $Ri^*_{min}=1$, $U_{max} k/N=1$; (h) $Ri^*_{min}=5$, $U_{max} k/N=1$; (i) $Ri^*_{min}=0.2$, $U_{max} k/N=2$; (j) $Ri^*_{min}=0.25$, $U_{max} k/N=2$; (k) $Ri^*_{min}=1$, $U_{max} k/N=2$; (l) $Ri^*_{min}=5$, $U_{max} k/N=2$. The case $Ri^*_{min}=0$ (for any $U_{max} k/N$) coincides with the black dotted lines.} 
\label{fig2}
\end{figure}

Examples of the bounds on $(\omega_r/N,\omega_i/N)$ in the complex plane are shown in figure \ref{fig2}, based on (\ref{semiell3}) (black lines) and (\ref{growth6}) (red lines), 
with $k U_{min}/N=0$.
Results are shown for three values of $\theta$ (solid, dashed and dotted lines), different values of $Ri^*_{min}$ (columns) and different values of $k U_{max}/N$ (rows). The case $Ri^*_{min}=0$ is omitted, since (\ref{growth6}) imposes no restriction on $\omega_i/N$ in that case, and condition (\ref{semiell3}) is entirely equivalent to that for $\theta=\pi/2$ at any value of $Ri^*_{min}$ (all black dotted lines in figure \ref{fig2}).

Since (\ref{growth6}) does not depend on $k U_{max}/N$, but only on $\theta$ and $Ri^*_{min}$, the red lines take the same values in each column of figure \ref{fig2}, namely for $\theta=0$, $\theta=\pi/20$ and $\theta=\pi/2$, respectively (i.e., solid, dashed and dotted red lines, respectively): $\omega_{imax}/N=0.5, 0.583, 1.118$, for $Ri^*_{min}=0.2$, $\omega_{imax}/N = 0, 0.396, 1$ for $Ri^*_{min}=0.25$, $\omega_{imax}/N= 0, 0.090, 0.5$ for $Ri^*_{min}=1$, and $\omega_{imax}/N=0, 0.036, 0.224$ for $Ri^*_{min}=5$. 

On the other hand, (\ref{semiell3}) prescribes a rescaling of the bounds limiting $(\omega_r/N,\omega_i/N)$ proportional to $k U_{max}/N$, as can be seen by comparing the various rows in figure \ref{fig2}, but is less sensitive to both $Ri^*_{min}$ and $\theta$, except when condition (\ref{condpos}) starts to apply. The variation of (\ref{semiell3}) with $\theta$, between $\theta=0$ and $\theta=\pi/2$, turns out to be faster for small $\theta$, hence results are displayed for an intermediate value of $\theta=\pi/20$ instead of, say, $\theta=\pi/4$. 

For $Ri^*_{min}=0.2$ or $Ri^*_{min}=0.25$, i.e., $Ri^*_{min} \le 0.25$, the bounds imposed on $(\omega_r/N,\omega_i/N)$ by (\ref{semiell3}) vary between those given by the semi-ellipse theorem ($\theta=0$) and by the semi-circle theorem ($\theta=\pi/2$). The maximum difference between the semi-ellipse and the semi-circle theorem is achieved for $Ri^*_{min}=0.25$ and $\theta=0$. While these bounds are unmodified by (\ref{growth6}) for relatively small values of $k U_{max}/N$ (see figure \ref{fig2}(a),(e)), (\ref{growth6}) restricts $\omega_i/N$ further as $k U_{max}/N$ increases (see figure \ref{fig2}(b),(f),(i),(j)). This is a consequence of the fact that, for constant $Ri^*_{min}$, the ratio of the spatial scale of the shear to that of unstable flow perturbations $k U_{max}/U'_{max}=k U_{max}/N Ri^{*1/2}_{min}$ becomes larger as $k U_{max}/N$ increases. 

It is worth noting that, for whatever value of $Ri^*_{min} \ge 0.25$, (\ref{growth6}) imposes $\omega_i/N=0$ for $\theta=0$, so the bounds prescribed by (\ref{semiell3}) become practically much less relevant in that case. For the same range of $Ri^*_{min}$, when $\theta < \pi/2$, (\ref{condpos}) becomes active, which makes the shape of the bounds imposed on $(\omega_r/N,\omega_i/N)$ by (\ref{semiell3}) differ markedly from a semi-ellipse or semi-circle, but rather be characterized by a flat cutoff corresponding to a constant value of $\omega_i/N$. For $Ri^*_{min}=1$ (figure \ref{fig2}(c),(g),(k)) this upper limit is $\omega_i/N=0.092$, and for $Ri^*_{min}=5$ (figure \ref{fig2}(d),(h),(l)) it is $\omega_i/N=0.036$. These limits are very close, but slightly higher than those prescribed by (\ref{growth6}) ($0.090$ and $0.036$, respectively), as noted above. Of course, when $\theta=\pi/2$, (\ref{growth6}) can still impose a substantially more restrictive maximum on $\omega_i/N$ than (\ref{semiell3}), as happens for $Ri^*_{min}=1$ or $Ri^*_{min}=5$ in figure \ref{fig2}(d),(h),(k),(l) (dotted lines).

\section{Conclusions}

We have analysed the constraints imposed by the integrated equations of motion on the inviscid instability of stratified flows with planar shear at an arbitrary angle to the vertical. Although Squire's theorem does not apply in general in this configuration, previous studies suggest that perturbations with 2D symmetry remain relevant, supporting the applicability of the present results. For example, those perturbations have been found by \cite{Deloncle_etal_2007} and \cite{Candelier_etal_2011} to be the most unstable for hyperbolic-tangent or Bickley jet shear profiles. 

The main contribution of this work is the derivation of integral constraints for the instability of planar shear flows that explicitly depend on the shear angle. These unify and extend classical results, recovering the Miles-Howard stability condition, and the semi-circle and semi-ellipse theorems as limiting cases for vertical and horizontal shear, respectively.

In particular, we show that no general stability condition exists for non-vertical shear (as is already known), but also derive a new upper bound for the instability growth rate as a function of the shear angle, consistent with numerical results by previous authors.
The extension of the semi-ellipse theorem, which is also new, leads to a more complicated bound for the complex angular frequency of unstable flow perturbations for intermediate shear angles, introducing also additional constraints on the growth rate.
These results provide a clearer analytical framework for understanding the instability of stratified planar shear flows with arbitrary orientation.







\begin{bmhead}[Acknowledgements.]
This work was funded by the Leverhulme Trust, under Grant RPG-2023-138. MACT acknowledges the Portuguese Funda\c{c}\~ao para a Ci\^encia e Tecnologia (FCT) for its financial support via the LAETA Associated Laboratory (project https://doi.org/10.54499/UID/50022/2025).
\end{bmhead}

\begin{bmhead}[Declaration of interests.]
The authors report no conflict of interest.
\end{bmhead}



\bibliographystyle{jfm}
\bibliography{jfm}






\end{document}